# Mars Thermospheric Variability Revealed by MAVEN EUVM Solar Occultations: Structure at Aphelion and Perihelion, and Response to EUV Forcing

**E. M. B. Thiemann[1], F. G. Eparvier[1], S. W. Bougher[2], M. Dominique[3], L. Andersson[1], Z. Girazian[4], M. D. Pilinski[1], B. Templeman[1] and B. M. Jakosky[1]**

[1]Laboratory for Atmospheric and Space Physics, University of Colorado, 1234 Innovation Dr., Boulder, CO 80303.
[2]Climate and Space Sciences and Engineering, University of Michigan, 2455 Hayward St, Ann Arbor, MI,  48109.
[3]Royal Observatory of Belgium, Avenue Circulaire 3, 1180 Uccle, Belgium.
[4]NASA Goddard Space Flight Center, Greenbelt, MD.

Corresponding author: Ed Thiemann (thiemann@lasp.colorado.edu)

**Key Points:**

- An extensive dataset of thermospheric neutral density and temperature is produced from MAVEN EUVM solar occultations.

- Altitude-latitude maps of thermospheric structure near aphelion reveal a temperature maximum near 0 LST, and minima near 3 and 20 LST in the northern hemisphere.

- A high altitude polar warming feature is observed at intermediate Northern latitudes near perihelion.

- Heating and cooling above and below 150 km, respectively, is observed that coincides with 27-day solar EUV variability.

- The average temperature sensitivity to ionizing EUV forcing is found to be $45 \pm 12$ K m$^2$ mW$^{-1}$ at the terminator, independent of dawn or dusk location.

## Abstract

The Mars thermosphere holds clues to the evolution of the Martian climate, and has practical implications for spacecraft visiting Mars, which often use it for aerobraking upon arrival, or for landers, which must pass through it. Nevertheless, it has been sparsely characterized, even when past accelerometer measurements and remote observations are taken into account. The Mars Atmosphere and Volatile EvolutioN (MAVEN) orbiter, which includes a number of instruments designed to characterize the thermosphere, has greatly expanded the available thermospheric observations. This paper presents new and unanticipated measurements of density and temperature profiles (120-200 km) derived from solar occultations using the MAVEN Extreme Ultraviolet (EUV) Monitor. These new measurements complement and expand MAVEN's intended thermospheric measurement capacity. In particular, because the local-time is inherently fixed to the terminator, solar occultations are ideally suited for characterizing long-term and latitudinal variability. Occultation measurements are made during approximately half of all orbits, resulting in thousands of new thermospheric profiles. The density retrieval method is presented in detail, including an uncertainty analysis. Altitude-latitude maps of thermospheric density and temperature at perihelion and aphelion are presented, revealing structures that have not been previously observed. Tracers of atmospheric dynamics are also observed, including a high altitude polar warming feature at intermediate latitudes, and an apparent thermostatic response to solar EUV heating during a solar rotation, which shows heating at high altitudes that is accompanied by cooling at lower altitudes.

## 1 Introduction

The Mars thermosphere (~100-200 km) couples the atmosphere with the near-space environment. It begins at the top of the well-mixed atmosphere at the temperature minimum known as the mesopause. 10-20 km higher lies the homopause [Bougher *et al.*, 2017a], where atmospheric species tend to diffusively separate according to mass. This results in (heavier) $CO_2$ being the dominant species below approximately 200-220 km, and (lighter) atomic O being the dominant species above 200-220 km [Nier *et al.*, 1977; Mahaffy et al., 2015a; Bougher et al., 2015b]. Near 170-200 km lies the exobase, where collisions become more rare. Above the exobase, atmospheric kinematics become less fluid-like and more particle-like, and the thermosphere blends into the exosphere, with no strict delineation between the two.

Solar extreme ultraviolet (EUV, 10-121 nm) radiation is the primary energy input to the thermosphere, with the majority of EUV irradiance being absorbed between 100 and 220 km, and the absorption peak occurring near 130 km. This results in a solar cycle dependence on thermospheric temperature, causing it to range from approximately $150^{o}K$ to $350^{o}K$ from solar minimum to maximum at the exobase [Bougher *et al.*, 2015a, Gonzalez-Galindo *et al.*, 2015]. In addition to EUV radiation, other factors, including energy transport from global circulation and upward propagating atmospheric waves, contribute to the thermospheric thermal balance [Bougher *et al.*, 2000; 2009; Medvedev *et al.*, 2011; 2015]. These factors can result in seemingly random fluctuations in thermospheric temperature and density, as can occur from gravity waves [e.g. Yiğit et al. 2015]; or quasi-periodic fluctuations as can occur from atmospheric tides [e.g. Lo *et al.*, 2015; Liu *et al.*, 2017]; or more gradual variations as can occur from large-scale circulation patterns occurring in the Mars upper atmosphere [e.g. Bougher *et al*., 2006]. $CO_2$ 15 um radiative cooling is significant in the lower thermosphere, near the EUV absorption peak, but is believed have a secondary influence on the overall thermal structure of the thermosphere [Bougher *et al*., 1999; Medvedev *et al*., 2015].

Understanding variability in the thermosphere's structure is important for characterizing global circulation, quantifying atmospheric escape and ultimately understanding the evolution of Mars's climate. More directly, the Martian thermosphere impacts spacecraft during their aero-braking campaigns, where atmospheric drag is used to dump momentum for orbit adjustment. For these reasons, the thermosphere has been characterized by a number of probes that have visited the red planet, beginning with remote measurements made by the first Mariner fly-by in 1965 [Fjeldbo *et al*., 1966], later by the Viking landers' in-situ measurements [Nier *et al.*, 1977] as well as during the aero-braking campaigns of the Mars Odyssey, Mars Global Surveyor and Mars Reconnaissance Orbiter missions [Keating *et al*., 1998; Withers, 2006; Bougher *et al*., 2017a], and most recently by the Mars Atmosphere and Volatile EvolutioN (MAVEN) mission [e.g. Mahaffy *et al*., 2015a; Bougher *et al*., 2015b; Evans *et al*., 2015; Jain *et al*., 2015; Elrod *et al*., 2016; Zurek *et al*., 2017]. Nevertheless, the Martian thermosphere is still poorly understood and observations of its variability continue to provide new insight into processes occurring in the Mars atmosphere. Understanding this variability and how it relates to the evolution of the Martian climate is one of the goals of MAVEN.

MAVEN is tasked with characterizing how the upper atmosphere of Mars varies in the current epoch in order to better understand how the atmosphere has evolved over time, and does so with a set of instruments that both characterize the solar drivers and the planet's response to these drivers [Jakosky *et al*., 2017]. MAVEN is able to characterize Mars's near-space environment and upper atmosphere with a highly elliptical 150 × 6000 km, 4.5 hour orbit, which allows MAVEN to make in-situ measurements of both the solar wind and upper atmosphere in the same orbit. The MAVEN orbit precesses such that in-situ measurements span a wide range of local-times and latitudes in the atmosphere in addition to the major regions of the near-space environment (e.g. solar wind, bow-shock, magnetosheath). MAVEN began science operations in November 2014 and has continued to operate nominally through the time of this writing. The current MAVEN mission spans over 1.5 Mars years, during which the solar cycle declined from moderate to minimum activity conditions [Lee *et al.,* 2017].

MAVEN includes three instruments that are designed to characterize thermospheric density: The Accelerometer (ACC) [Zurek *et al*., 2015], the Neutral Gas and Ion Mass Spectrometer (NGIMS) [Mahaffy *et al*., 2015b], and the Imaging Ultraviolet Spectrograph (IUVS) [McClintock *et al*., 2015]. ACC and NGIMS make routine in-situ density measurements near periapsis, with typical reliable measurement sensitivity between 145 and 250 km for NGIMS, and between 145 and 170 km for ACC. Note, the ACC and NGIMS minimum measurement altitudes are lowered to near 120-130 km during periodic "Deep Dip" Campaigns [Bougher *et al*., 2015b; Zurek *et al*. 2017], corresponding to a periapsis density corridor that is raised to approximately 2-3.5 kg/km³ for 1 week intervals. These campaigns have occurred at a cadence of approximately once every three months. IUVS measures thermospheric density remotely through routine limb scan measurements of atmospheric airglow [Evans *et al*., 2015; Jain *et al*., 2015]. These observations are made with multiple scans during the periapsis orbit segment, observing locations ranging from approximately 100 to 1000 km perpendicularly away from the orbit track, with a reliable measurement sensitivity between 120 and 180 km altitude. Additionally, IUVS measures thermospheric density during special stellar occultation campaigns [Gröller *et al.*, 2015], occurring approximately every 2 months. The IUVS stellar occultation observations measure atmospheric density from near the surface to 150 km.

This paper presents new thermospheric density measurements between 120 and 200 km, retrieved from solar occultation measurements made by the MAVEN EUV Monitor (EUVM) [Eparvier *et al.* 2015]. The primary purpose of MAVEN EUVM is to characterize solar EUV irradiance at Mars. This paper demonstrates that thermospheric densities can be accurately inferred from the extinction of the solar signal measured by EUVM as the spacecraft enters or exits eclipse since EUV radiation is absorbed entirely in the thermosphere. Because solar occultations are inherently confined to the terminator, the EUVM solar occultations (EUVM-SOs) are generally not collocated with those made by the other MAVEN instruments, broadening the global coverage of neutral density measurements made by MAVEN during a single orbit. Furthermore, the fact that the local time (LT, not to be confused with *Local Solar Time*, which depends on the solar declination angle and varies at the terminator with latitude) is fixed to either 06:00 (at the dawn terminator) or 18:00 (at the dusk terminator) make EUVM-SOs more suitable for characterizing long-term or latitudinal variability than the other MAVEN neutral density measurements, which must contend with varying LT and/or solar zenith angle. This paper begins by describing the method used to retrieve density from the EUVM-SOs. The data used in the density retrievals are then presented, followed by the estimated uncertainty associated with the density retrievals. Next, results of the density profiles made during the MAVEN mission to-date are presented. These results include altitude-latitude maps of density and temperature, as well as an analysis of how thermospheric temperature varies with solar EUV irradiance and season. These results are discussed and put into the context of our current understanding of the Mars atmosphere and existing global model simulations.

## 2 Methods

### 2.1 Retrieving Neutral Density from EUVM Solar Occultations

The method used to retrieve thermospheric neutral density from the MAVEN EUVM measurements is based on a solar occultation retrieval method first developed by Roble and Norton [1972], who used full-disk irradiance measurements near 121 nm to characterize thermospheric $O_2$ at Earth. This method was updated by Thiemann *et al.* [2017a], who used full-disk irradiance measurements between 0.1 and 20 nm to characterize thermospheric $N_2$ and O, also at Earth. A brief description of the retrieval method is summarized here, but the reader is referred to those two papers for a detailed description of the retrieval method. In particular, the study of Thiemann *et al.* [2017a] was motivated by an effort to validate methods used in this study by applying them at Earth, where thermospheric densities are better understood. As such, the method of Thiemann *et al.* [2017a] is nearly identical to the method used in this study. Deviations between the method used here and that by Thiemann *et al.* [2017a] are specified as they arise.

The Extinction Ratio (ER) relates the observed solar intensity, *I*, at tangent height, $h_t$, to that at the top of the atmosphere ($I_\infty$) according to

$$\mathbf{ER} = \frac{I(h_t)}{I_\infty} = \exp(-\sum_i N_i(h_t)\,\sigma_i), \quad (1)$$

where $N_i(h_t)$ and $\sigma_i$ are the column density along the line-of-sight and absorption cross-section of the $i^{th}$ absorbing species, respectively. $h_t$ is the height above the surface tangent to the line-of-sight. Equation (1) is forward modeled to find the altitude profile of $N_i$ that results in a predicted ER in closest agreement with a measured ER. Full-disk solar EUV irradiance measurements, such as those made by MAVEN EUVM, present additional complications because the spatial extent of the solar disk at the limb is comparable to the atmospheric scale-height, and EUV radiance varies significantly over the solar disk. These complications are addressed by incorporating solar images and reference atmospheres into the retrieval algorithm. Specifically, signal extinction is forward modeled by integrating the intensity of the solar disk propagating through a reference atmosphere when predicting $N_i$ at a particular $h_t$. The process is iterated until the scale-height between 150 and 180 km of the reference and retrieved atmospheres agree to within 5%.

The retrievals presented here assume a pure $CO_2$ atmosphere, allowing the summation in Equation (1) to reduce to $N_{CO2}\sigma_{CO2}$. This assumption is justified for the following reasons: There typically is minimal extinction of the EUVM signal at altitudes where O becomes the major species. Further, the $CO_2$ cross-section is 3-5 times larger than that of O in the 17-22 nm wavelength range, where EUVM-SO measurements are made. As such, at 200 km, where the O to $CO_2$ ratio is close to 0.25 [e.g. Bougher *et al.* 2015b], the contribution of O to the observed extinction is less than 10%. Nevertheless, neglecting O introduces measurable error at high altitudes, and evidence of O contaminating the retrieved density profiles is evident in some cases. Therefore, caution is needed when analyzing data at or above 200 km, and any scale-height increases at these altitudes should be treated with suspicion because the observed temperature is expected to be constant above approximately 170 km, where atmospheric collisions become negligible. This assumption of a pure $CO_2$ atmosphere is accounted for in the uncertainty estimates presented in Section 4.

Although the Mars upper atmosphere can be approximated as having two scale-heights between 100 and 200 km, one corresponding with the colder well-mixed lower atmosphere and the other with the warmer thermosphere, which warms rapidly above the mesopause resulting in an inflection near 120 km on a linear-logarithmic scale, only single scale-height atmospheres are used as the reference atmospheres. The reference atmospheres used in the retrieval are isothermal with the form $n(z)=n(z_0)\exp((z-z_0)/H)$, where $n$, $z$, $z_0$ and $H$ are the number density, altitude, reference altitude and scale-height, respectively. Roble and Norton [1972] and Thiemann *et al.* [2017a] showed that isothermal reference atmospheres do not force the retrieved atmospheres to also be isothermal, and temperature gradients are accurately captured at altitudes above an abrupt change in scale-height. However, error induced by the reference atmospheres increases markedly with decreasing altitude below an abrupt change in scale-height. The error introduced due to using single scale-height reference atmospheres is quantified in the uncertainty estimates presented in Section 4.

Once Equation (1) has been solved for $N_{CO2}(h_t)$ for every $h_t$ in a given occultation scan, Abel inversion is used to convert $N_{CO2}(h_t)$ to $n_{CO2}(h_t)$, the number density at each $h_t$. The Abel inversion requires the assumption of spherical symmetry, which is questionable at the terminator. However, this assumption is reasonable when considering it is required only over the ~400 km column of atmosphere sampled in the retrieval, which corresponds with approximately $6.8^o$ of longitude. Further, Thiemann *et al.* [2017a] compared Earth measured thermospheric density retrievals with climate model predictions and found no systematic biases due to the spherical symmetry assumption.

Figure 1 shows sample data and fits from orbits 439 and 440, occurring on 21 December 2014, in order to demonstrate the data quality in the intermediate steps of the retrieval process. Figures 1a and 1b show the measured ER with black curves. Note, the typical EUVM measurement noise is 30 ppm [Thiemann, 2016] and neither it nor solar variability are distinguishable in typical ER measurements. The effects of both types of variability on the retrieval uncertainty are discussed in Section 4. For each measured ER value, $N_{CO2}$ is found by forward modeling Equation (1); this result is shown with the red-dashed curves. Because the numerical solver can fail to find solutions at the ER extremes, only ER measurements in the 0.01-0.99 range are used. The fractional model-measurement difference is shown with the blue curve and right-hand axis, and is less than 1 ppm. Figures 1c and 1d show the corresponding retrieved $N_{CO2}$ profiles with black curves, which are inverted for number density using an Abel transform. Because the Abel integral extends to infinity, the topside column density is extrapolated to 3000 km altitude (essentially infinity) by fitting the upper quartile $N_{CO2}$ values to an exponential function and applying the fit results to altitudes above the highest retrieved $N_{CO2}$ value through 3000 km; this profile is shown with the red-dashed curves.

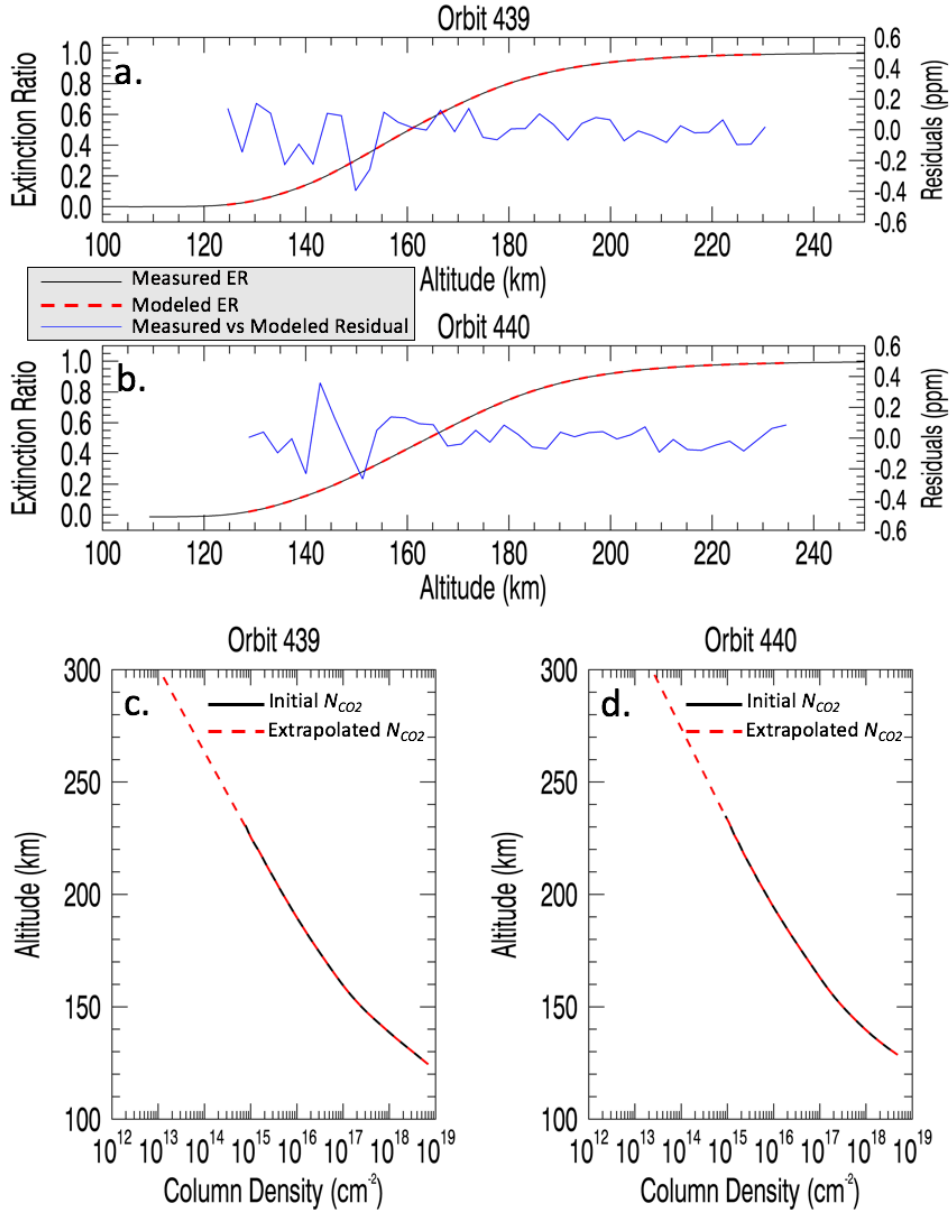

**Figure 1.** Panels a and b show sample extinction ratios with black curves. The resulting ER fits from solving Equation (1) are shown with red dashed curves. Panels c and d show the retrieved column densities and the extrapolated column densities, the latter of which is inverted for number density, with black and red-dashed curves, respectively.

### 2.2 Method for Inferring Temperature from Density Profiles

Once $n_{CO2}(h_t)$ has been retrieved, a corresponding temperature profile, $T(h_t)$, is inferred using the method of Snowden *et al.* [2013]. This method infers temperature by first computing pressure, $p(h_t)$, at each altitude by integrating $n_{CO2}(h_t)$ above each altitude to infinity. Once the pressure is found, hydrostatic equilibrium is invoked and $T(h_t)$ is found from the ideal gas law,

$$T(h_t) = \frac{p(h_t)}{n_{CO2}(h_t)k} \,, \quad (2)$$

where $k$ is Boltzmann's constant. The method used here differs from that of Snowden *et al.* [2013] in that in this work $n_{CO2}(h_t)$ is extrapolated to high altitudes (essentially infinity) whereas Snowden *et al.*, [2013] estimated the absolute pressure at the top of the density profile. Density extrapolation to high altitudes is possible here because the EUVM-SO measurements retrieve density above the exobase, where the atmosphere is approximately isothermal. Isothermality allows for the density profile above the exobase to be approximated by

$$n(h_t > z_{exo}) \cong n(z_{exo})\exp(-(h_t - z_{exo})\frac{m_{CO2}g}{kT_{exo}}), \quad (3)$$

where $z_{exo}$ is the exobase altitude, $m_{CO2}$ is the $CO_2$ molecular mass, $g$ is the local gravity and $T_{exo}$ is the exospheric temperature.

In practice, $T_{exo}$ is found by fitting the measured $n_{CO2}(h_t)$ between $5\times10^7$ cm$^{-3}$ and $2\times10^8$ cm$^{-3}$ to Equation (3), where the lower density bound corresponds with where the measurement uncertainty becomes large and the upper density bound corresponds with where the mean-free-path of a $CO_2$ molecule is approximately equal to a scale height, which is the traditional definition of the exobase. Equation (3) is then used with $T_{exo}$ to extend the measured $n_{CO2}(h_t)$ 100 km above where $n_{CO2}(h_t) = 5\times10^7$ cm$^{-3}$, which is essentially infinity for the purposes of calculating pressure relative to altitudes where EUVM-SO measurements are made. $p(h_t)$ is then found by solving,

$$p(h_t) = m_{CO2}g \int_{h_t}^{\infty} n_{CO2}(h_t)dz. \quad (4)$$

## 3 Data Used for Retrieving Density

The EUVM-SO density retrievals require the following data as inputs: EUVM solar measurements, the EUVM instrument response function, a contemporary solar irradiance spectrum, $CO_2$ absorption cross-sections and a contemporary solar image. Each of these inputs will be described in this section.

MAVEN EUVM's primary mission is to characterize the solar EUV input into the Mars upper atmosphere and does so using three photometer channels that measure full disk solar EUV irradiance over different wavelength ranges. EUVM uses a simple optical design, with each channel consisting of an entrance-aperture, a Field of View (FOV) limiting aperture, a normal incidence filter and an AXUV technology [Gullikson *et al.*, 1996] Si detector. The channel band-passes are 0.1-7 nm, 0.1-7 nm + 17-22 nm, and 117-127 nm, determined by an Al/Ti/C foil filter, an Al/Nb/C foil filter, and an Al on MgF$_2$ substrate interference filter, respectively. The three band passes were chosen because they originate in distinct regions of the Sun's atmosphere, each with a distinct characteristic variability, providing a diverse set of inputs to the Flare Irradiance Spectral Model-Mars (FISM-M) [Thiemann *et al.* 2017b], which predicts solar spectral irradiance between 0.1 and 190 nm. The direct irradiance measurements and FISM-M modeled irradiances are reported in the MAVEN EUVM Level 2 and Level 3 data products, respectively, publicly available through the NASA Planetary Data System (PDS).

The EUVM-SO densities reported here are from the Al/Nb/C channel of the EUVM instrument, which nominally measures irradiance in the 17-22 nm band but includes a pass-band in the 0.1-7 nm range. The fully-calibrated irradiances are not used for EUVM-SO measurements because they incorporate a spectral model to isolate and remove the 0.1-7 nm contribution from the measurement, and this contribution must be considered in the EUVM-SO measurements because it contributes to the observed extinction profile. Instead, partially-calibrated irradiances are used that have been corrected for dark current, a visible light leak, temperature, gain drifts, and FOV effects. These partially calibrated irradiances are stored in the "Corrected Counts" variable of the MAVEN EUVM Level 2 data product.

The response function of an EUVM channel determines which wavelengths of incident radiation contribute to the measured signal, and depends on the filter transmission and the responsivity of the detector for each channel. Quantitatively, it is the photocurrent produced per incident optical power as a function of wavelength. Figure 2 shows the modeled response function and uncertainty for the Al/Nb/C channel. The EUVM channel response functions were characterized at the National Institute of Standards and Technology (NIST) Synchrotron Ultraviolet Radiation Facility (SURF) III [Arp *et al.* 2000], and detailed results of the characterization are reported in Thiemann [2016]. SURF III produces spectra at EUV wavelengths whose intensity and spectrum are strictly constrained by the synchrotron beam-current and energy, both of which are well characterized, resulting in spectral uncertainties smaller than 0.6% at wavelengths above 1 nm [Arp *et al.* 2000]. Multiple beam energies, each with a unique characteristic spectrum, were used to probe each EUVM channel response function, and the detector and filter thicknesses used in the response function model were adjusted until the instrument response (i.e. the ratio of input optical power to output signal) was constant across all beam energies. This method was used by Woods *et al.* [1999; 2005] to calibrate the soft x-ray and EUV photometers on the X-Ray Photometer System (XPS) onboard the Solar Radiation and Climate Experiment (SORCE) and the Thermosphere Ionosphere Mesosphere Energetics and Dynamics (TIMED) missions. Thiemann [2016] updated this method to account for defects in the foil filters, termed pin-holes, that allow light to pass directly through the foil as the name suggests [Powell, 1993]. Thiemann [2016] found that the characterized response function filter and detector thickness are in substantially better agreement with those reported by the manufacturer when pin-holes are considered and, therefore, the response function model that includes pin-holes is considered to be the best estimate. Because of the indirect nature of the response function characterization, its associated uncertainty is difficult to quantify. For the purposes of computing the density retrieval uncertainty estimate presented in Section 4, the response function uncertainty is conservatively estimated as the difference between the response function model that includes pin-holes and the model that does not, divided by $\sqrt{2}$.

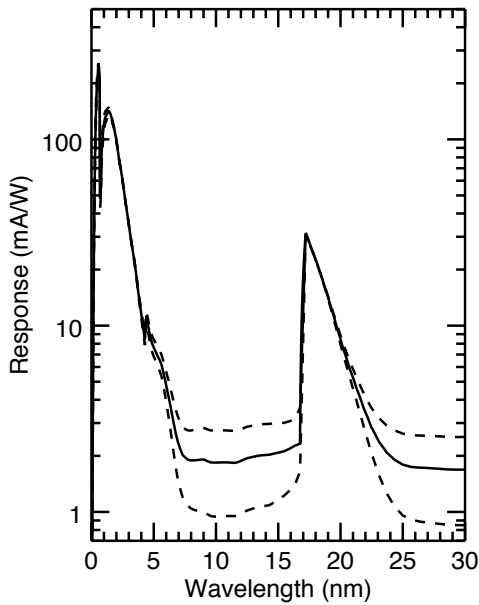

**Figure 2.** The response function (solid curve) and uncertainty (dashed-curves) for the EUVM Al/Nb/C channel.

Daily averaged FISM-M solar spectra are used in the density retrievals. The bin-to-bin solar spectral uncertainty over the Al/Nb/C channel pass-band is approximately 30% in the 0.1-7 nm range and 5% in the 17-22 nm range. The FISM-M spectra and uncertainties are described in detail in Thiemann et al. [2017b]. EUV spectral irradiance can vary rapidly and significantly during solar flares [e.g. Woods *et al.*, 2011; Thiemann *et al.*, 2017b; 2017c]; therefore, occultations occurring during flares are neglected. Note that the absolute uncertainty in spectral irradiance does not matter for solar occultations since the measurements are normalized by their value at the top of the atmosphere.

The $CO_2$ cross-sections used are from Huestis and Berkowitz [2011] and correspond with a temperature of 300 K.

The radiance over the solar disk in the Al/Nb/C channel bandpass is approximated as being the same as that in the bandpass of the 17.4 nm SWAP (Sun Watcher using Active Pixel system and image processing) telescope [Berghmans *et al.* 2006; Seaton *et al.* 2013] onboard the Project for OnBoard Autonomy 2 (PROBA2) Earth orbiting satellite. This approximation is valid because emissions measured in both instruments' band-passes are dominated by plasmas forming at similar temperatures [Berghmans *et al.,* 2006, Thiemann *et al,.* 2017b]. Earth measured images are interpolated to Mars for a given day by time-shifting earlier and later images from Earth using the irradiance interpolation method described in Thiemann *et al.* [2017b], which interpolates irradiance assuming it varies on solar rotation time-scales. Images used are measured nearest to 12:00 UT that are also at least 2 hours after a C-class or larger solar flare.

The image interpolation method has some error associated with it because the radiance on the solar disk also evolves due to processes occurring on time-scales shorter than that of solar rotation [e.g. active region emergence and decay]. This error is estimated by comparing SWAP images measured on a given day, say, $t_d$, with predicted images found by averaging SWAP images from one solar rotation prior (day $t_d$ - *27*) and one solar rotation later (day $t_d$ + *27*). Since Equation (1) inherently assumes the Mars thermosphere is horizontally stratified, the solar disk is inherently integrated in the direction parallel to the Martian surface. Figure 3 shows this uncertainty as a function of the subtended angle from disk center, along an axis in the zenith direction that bisects the solar disk. Note the solar corona extends beyond the visible disk, which subtends an angle of $0.52^o$ at 1 AU.

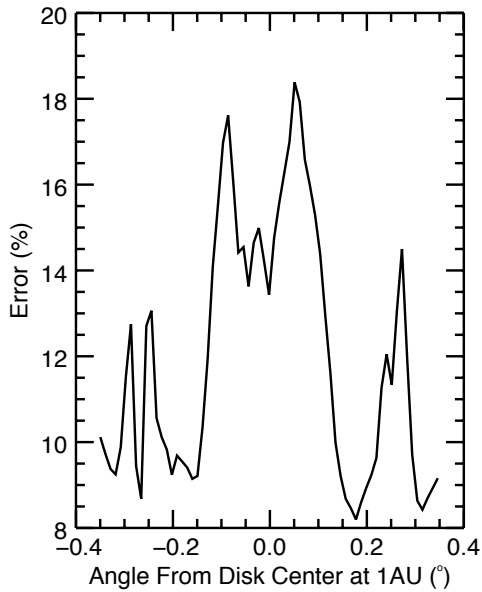

**Figure 3.** Error associated with interpolating EUV images measured at Earth to Mars. Error is shown as a function of angle along the zenith direction from the Martian surface projected onto the solar disk.

## 4 Density Retrieval Uncertainty

The EUVM-SO density uncertainty and systematic error are estimated with Monte-Carlo (MC) analysis. Ad-hoc "ground-truth" atmospheres are synthesized over the range of densities observed in the EUVM-SO dataset and compositions (only $CO_2$ and O are considered) expected to occur in the Mars thermosphere. The $CO_2$ density profiles are synthesized by adding two decaying exponential functions of the form $n(z) \cong n(z_o) \exp(-(z - z_o)\frac{m_{CO2}g}{kT})$ with $z_0$ equal to 100 km and 150 km. $n(z_0)$ for the lower (upper) exponential term is $5\times10^{12}$ ($5\times10^9$) cm$^{-3}$ $\pm$ 50% (1-sigma) with a lower limit of $10^{10}$ ($10^7$) cm$^{-3}$. T for the lower exponential term is $100^o$ K $\pm$ 80% with a lower limit of $70^o$ K, and T for the upper exponential term is 2.5× that of the lower exponential term with an additional 50% randomization and a lower limit of 150 K. The O density profiles are synthesized with a single exponential function of the same form as that used for $CO_2$, with O density fixed to be 1% of the $CO_2$ density at 150 km and the same temperature as the upper exponential $CO_2$ density term. Figures 4b and 4c show the mean $CO_2$ density and O/$CO_2$ density ratio with solid curves and the corresponding $\pm$ 1-sigma range with dashed curves.

The ground-truth densities are then input to a measurement model of the Al/Nb/C channel to synthesize extinction ratios. Note, O photoabsorption cross-sections are from PHoto Ionization/Dissociation Rates (PHIDRATES) database [Huebner and Mukherjee, 2015]. Densities are then retrieved from these extinction ratios and compared to the ground-truth densities to quantify the random uncertainty and systematic error in the measurement.

The uncertainty analysis includes the following sources of random uncertainty or systematic error: measurement noise, uncertainty in the instrument response function, non-flaring solar variability, solar spectral uncertainty, $CO_2$ cross section uncertainty, image interpolation uncertainty and uncertainty in the retrieval algorithms. The measurement noise is very small and approximately 30 ppm of the measured signal [Thiemann 2016]. The $CO_2$ cross section uncertainty at the measurement wavelengths is 5% [Gallagher *et al*., 1988]. The typical non-flaring solar variability estimate is taken from Thiemann et al. [2017a], who quantified the typical non-flaring EUV variability below 20 nm over solar occultation time-scales and found it to be 0.35% over a 2-minute occultation scan. All other uncertainties are discussed in Section 3, except for that of the retrieval algorithm itself, which is inherently characterized by the MC analysis.

Uncertainty in the cross section and response function result in systematic errors in the density retrieval that depends on the sign of the respective uncertainty (i.e. whether the response function or cross sections are actually smaller or larger than estimated). Because of this, the systematic error is simulated for two cases, spanning the possible range the summed systematic errors. The total systematic error is bounded by both cases.

Figure 4a shows the results of the MC uncertainty analysis over the density range that the EUVM-SO measurements are sensitive. On the right-hand side of the figure, approximate corresponding altitudes at perihelion and aphelion are shown. The solid curve corresponds with the random uncertainty and is due to signal strength, instrument noise, solar variability, and uncertainty in the solar spectrum. The random uncertainty is between 5 and 10 % over much of the density range, but becomes larger at the highest and lowest densities. This is where the extinction ratio is close to 0 or 1, respectively, and the change in the extinction ratio with altitude near these extremes becomes comparable to the magnitude of random variations in the signal. The dashed curves are the envelope that bounds the systematic error. As discussed above, the systematic error is ill defined because of the uncertainty in the response function. At intermediate densities and altitudes, the systematic error is somewhere between approximately 5 and 22%. At low densities and high altitudes, the systematic error becomes large due to extinction from unaccounted O, which becomes increasingly substantial with increasing altitude. At high densities and low altitudes, the systematic error becomes large due to the assumption of a single-scale height atmosphere, when a second scale-height corresponding with the cold mesopause dominates at lower altitudes.

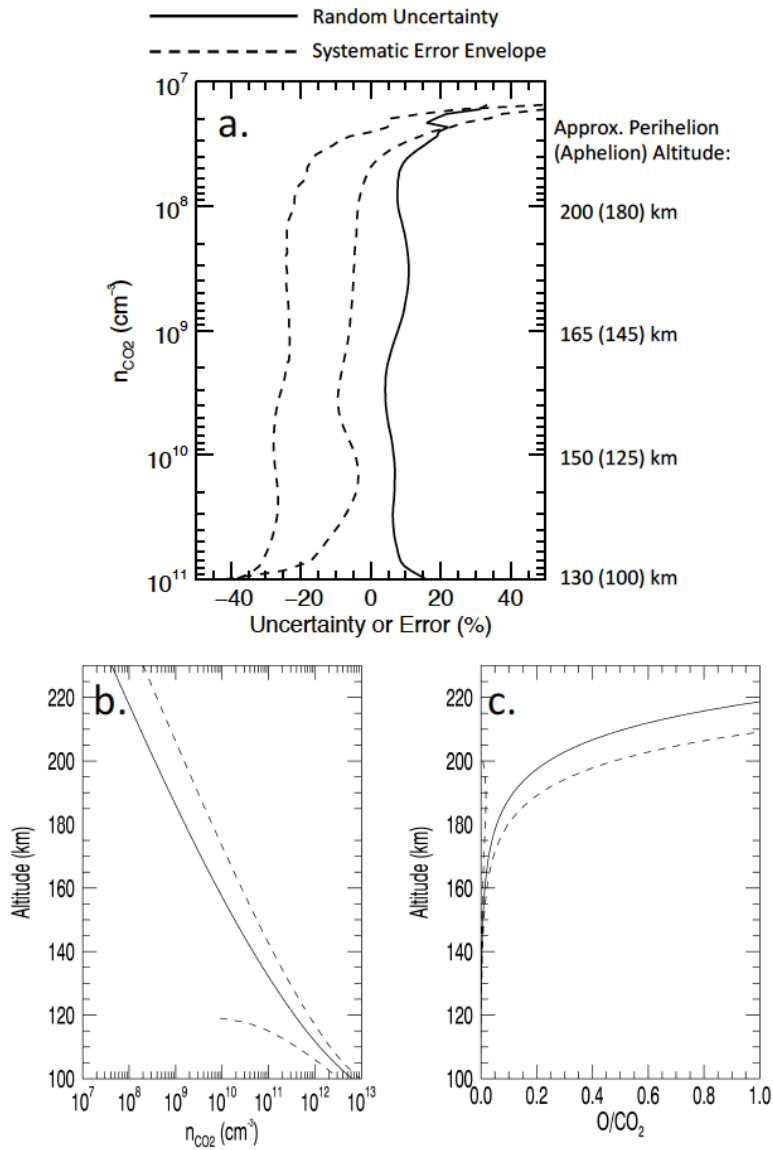

**Figure 4.** (a) EUVM-SO random uncertainty (solid curve) and systematic error (dashed curves) versus $CO_2$ number density. The systematic error is bounded by the two dashed curves shown, and is ill-defined because the sign of any systematic error in the response function is unknown. On the right-hand side, approximate altitudes corresponding with density at perihelion (aphelion) are shown. The average and standard deviation for the ground-truth density profiles (b) and $O/CO_2$ density ratios (c) are shown with solid and dashed curves, respectively.

The random uncertainty and systematic error have implications for how the EUVM-SO profiles are interpreted. Clearly, data at the altitude and density extremes need to be treated with caution. For intermediate altitudes, the systematic error could be removed from the data to improve accuracy using the profile(s) shown in Figure 4. This is not done for the data presented in Section 5 because doing so would not materially change the conclusions. Instead, the estimated systematic error is conservatively treated as an additional uncertainty. The small random uncertainty across the majority of the sensitivity range indicates that the EUVM-SO data are ideal for tracking relative changes. In particular, the small random uncertainty in the density profiles indicate relatively small temperature changes are resolvable given the logarithmic dependence of temperature on density for an atmosphere in hydrostatic equilibrium.

Thiemann et al. [2017a] performed a nearly identical MC uncertainty analysis for the LYRA solar occultation measurements at Earth. The two differences in their analysis are that they omitted the response function uncertainty because it was known to remain within 1% due to a more precise pre-launch calibration, and they had an additional systematic uncertainty associated with assuming equivalent cross-sections for $N_2$ and O. Nevertheless, the uncertainties for LYRA are comparable to those shown for EUVM in Figure 4. Further, Thiemann *et al.* [2017a] found better than 5% agreement with the LYRA measurements and predictions made by the NRLMSISE-00 empirical model for Earth's thermospheric density [Picone *et al.*, 2002], which has been independently characterized and validated by available thermospheric density measurements at Earth. The good agreement between LYRA solar occultation observations and NRLMSISE-00 predictions also serve to validate the EUVM-SO observations given the similarities in the instruments and near identical retrieval algorithms.

Additionally, if the observing latitude varies substantially during an occultation scan, as occurs when the spacecraft β-angle approaches 90°, the retrieved densities no longer correspond with a vertical profile. This has no significant effect on the density retrieval [Thiemann *et al.*, 2017a] but, because the temperature calculation assumes a vertical density profile, substantial error can be introduced to derived temperatures when the scan spans more than ~4° latitude.

## 5 Results

At the time of this writing, over two thousand density profiles have been retrieved from EUVM-SO measurements, providing a unique dataset for studying the structure and variability of the Mars thermosphere. It would be unwieldly to present the dataset in its entirety in a single paper. As such, the focus of this section is to present a subset of the observations that provide new insight into the structure of the Mars thermosphere and its variability, while highlighting the utility and novelty of the dataset. Specifically, this section presents measurements of: i) the temperature and density structure of the northern hemisphere near aphelion at both dawn and dusk terminators; ii) the temperature and density structure in the dusk terminator near perihelion; and iii) the correlation between thermospheric temperature variability and solar variability.

In general, solar EUV heating, planetary rotation, and tidal, planetary and gravity wave forcing from below determine the density and temperature structure of the thermosphere, resulting in regional temperature and density enhancements or depletions [Bougher, 1995, Bougher et al., 2002; 2015c]. The thermospheric structure has been studied using a number of Mars specific General Circulation Models (GCMs), including the Mars Thermosphere GCM (MTGCM) [Bougher *et al.*, 1999; 2000], the Laboratoire de Météorologie Dynamique Mars GCM (LMD-MGCM) [Gonzalez-Galindo *et al.*, 2009a; 2009b], and the Mars Global Ionosphere Thermosphere Model (M-GITM) [Bougher *et al.*, 2015a]. Throughout the following data presentation, GCM results are referenced to aid the interpretation of features observed in the EUVM-SO data. These comparisons are predominantly qualitative, and hindered by the lack of previously published scenarios that exactly correspond with that observed by the EUVM-SOs. A more complete understanding of the features in the EUVM-SO data can be gained by direct GCM to measurement comparisons, and will be the topic of future studies.

Figure 5 provides an overview of the EUVM measurements made and processed at the time of this writing. Figure 5a shows inferred thermospheric temperature at 585 nPa ($T_{595}$) (approximately 150 km at perihelion) versus time at the dawn and dusk terminators with blue and red asterisks, respectively; also shown in this figure is the inverse Mars-Sun distance squared ($1/R^2$) with green asterisks, which is proportional to solar insolation, and the corresponding 0-93 nm, $CO_2$ ionizing EUV irradiance ($E_{EUV}$) with black asterisks, which is derived from the EUVM Level 3 data product (Thiemann *et al*. 2017b). Comparing the $T_{595}$ measurements with the solar values reveals a clear correlation between them, which will be investigated further in Section 5.3. Figure 5b shows the latitudes corresponding with the temperature measurements using the same color as in Figure 5a. The MAVEN orbit precession results in EUVM-SO measurements being made over a range of latitudes within a relatively short time. As a result, EUVM is capable of scanning a wide range of latitudes during which the seasonal changes are relatively small, enabling the profiling of entire hemispheres at fixed local time and approximately fixed season. This capability is exploited next in Sections 5.1 and 5.2, where maps of thermospheric structure near aphelion and perihelion are presented.

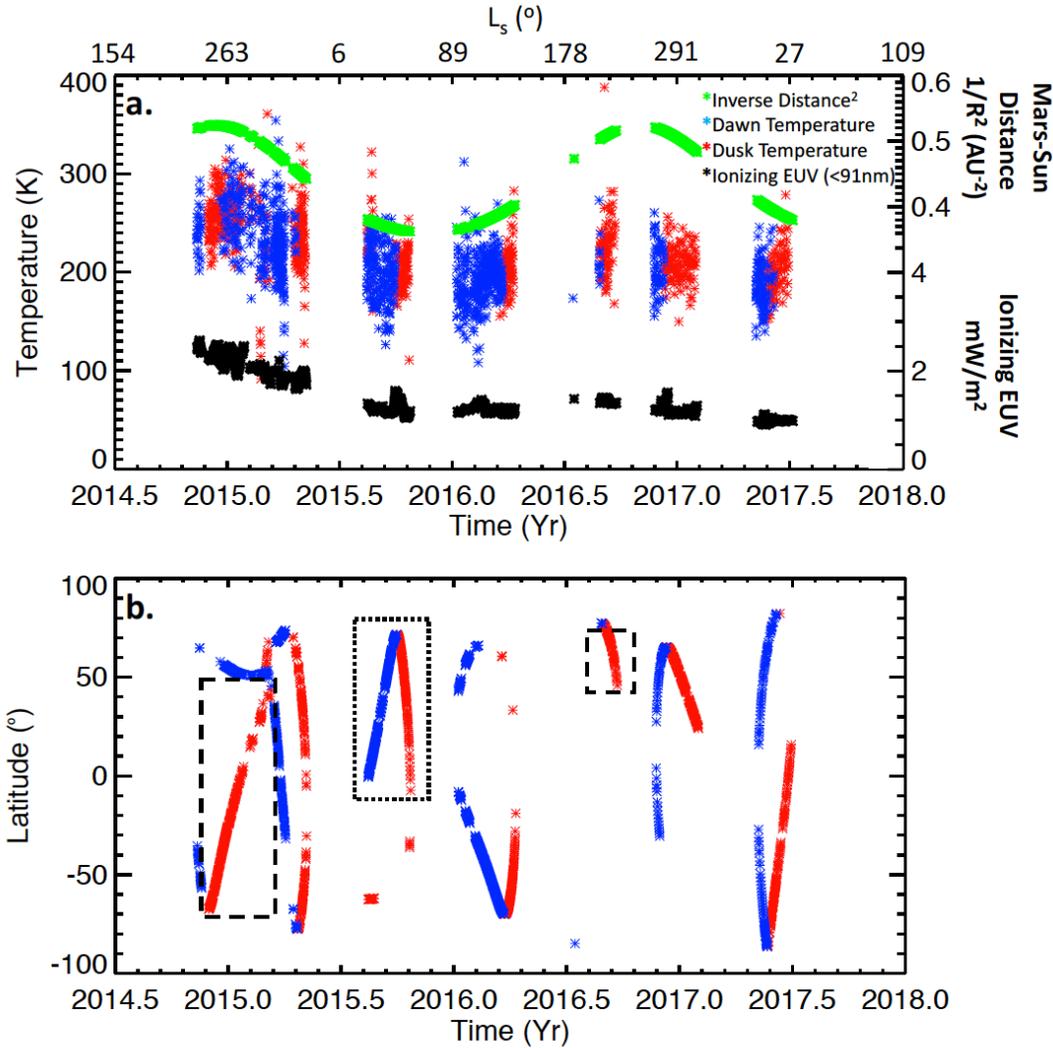

**Figure 5**. Thermospheric temperature variability observed by MAVEN EUVM-SO measurements. a) Dawn and dusk temperatures at 595 nPa are shown with blue and red asterisks, respectively. Also shown are measures of solar variability, with the green asterisks showing the inverse Mars-Sun distance squared, $I/R^2$, and the black asterisks showing $CO_2$ ionizing EUV radiation, $E_{EUV}$. b) Latitudes of where the EUVM observations are made, where blue and red asterisks again correspond with measurements at the dawn and dusk terminators, respectively. The data in the dotted box are analyzed in Section 5.2 and the data in the dashed boxes are analyzed in Section 5.1.

## 5.1 Mars Thermospheric Structure near Aphelion: Northern Dawn and Dusk Terminators

Between early August and early November of 2015, EUVM-SOs scanned much of the Northern Hemisphere, beginning near the Dawn Equator, moving over the North Pole, and finishing near the Dusk Equator. During this time, the Mars solar longitude, $L_s$, varied from 23.2° to 64.5°, corresponding with mid-Spring in the Northern Hemisphere, as the planet approached aphelion (occurring at Ls=71°). The non-zero solar declination during this time resulted in EUVM-SOs observing a maximum latitude of 71.6°N as the observations passed behind the pole at the Dawn-Dusk transition.

Figure 6 shows EUVM-SO measured temperatures and densities from this time period, where panels a (c) and b (d) show temperature (density) as a function of latitude versus altitude or pressure, respectively. The bottom horizontal axis shows the corresponding latitude. Ambiguity in dusk versus dawn latitudes is resolved with the top horizontal axis, which gives the corresponding local solar time. Panels e and f show density and temperature, respectively, at 150 km, where black curves correspond with individual orbits and purple with the 5-orbit running mean. Note, the mean latitude is used for each occultation scan, resulting in the density or temperature from each occultation scan being (artificially) rendered as perfectly vertical in Figure 6. The overlaid hatch pattern in Figures 6a and 6b corresponds with scans where the latitude varies by 4° or more, implying the derived temperatures are prone to error as discussed in Section 4.

The temperature and density maps shown in Figure 6 reveal similarities and differences between the dawn and dusk terminators. The temperature magnitude and variation with altitude shown in Figure 6a are in general agreement with GCM predictions at 04:00 and 15:00 LST by Bougher *et al.*, 1999. Examining the highest altitudes where measurements are available in Figure 6c reveals a substantial 20-30 km dusk bulge, which can be attributed to diurnal variability to first-order because, prior to observation, the dusk atmosphere is rotating from the solar EUV heated day-side, while the dawn atmosphere is rotating from the cold night-side [Bougher *et al.*, 1999]. Additionally, the dawn terminator is more variable than the dusk terminator as can be seen in Figures 6a and 6b by a marked increase of high frequency structure at high altitudes at the dawn terminator; this is consistent with recent observations reported in Zurek *et al.*, 2017 showing larger mass density variability at dawn than dusk below 150 km. This high frequency variability may be more temporal than latitudinal in nature given that these maps are constructed over many orbits. Simulations of the Mars thermospheric structure near the equinoxes indicate a much steeper temperature gradient at dawn relative to dusk [Bougher *et al.*, 1999; Gonzales-Galindo *et al.*, 2009a, Bougher *et al.*, 2015a], which may partially explain the increased temperature variability observed at dawn.

Specifically, this temporal variability may be due to gravity waves, which are expected to have a greater influence on thermospheric temperature when vertical temperature gradients and, hence, kinematic viscosity gradients increase [Vadas *et al.*, 2014]. Enhanced gravity wave activity near the dawn terminator is supported by simulations [Yigit *et al.*, 2015a] and recent observations [Terada *et al.*, 2017]. Figure 6d shows increased orbit-to-orbit density variability clustered around ~55°N at both terminators. These same latitudes correspond with a 20-30° K drop in temperature relative to that at 71.6°N, and may be evidence of high latitude gravity wave induced cooling predicted by some simulations [Medvedev & Yiğit, 2012; Medvedev *et al.*, 2015; Yiğit *et al.*, 2015b].

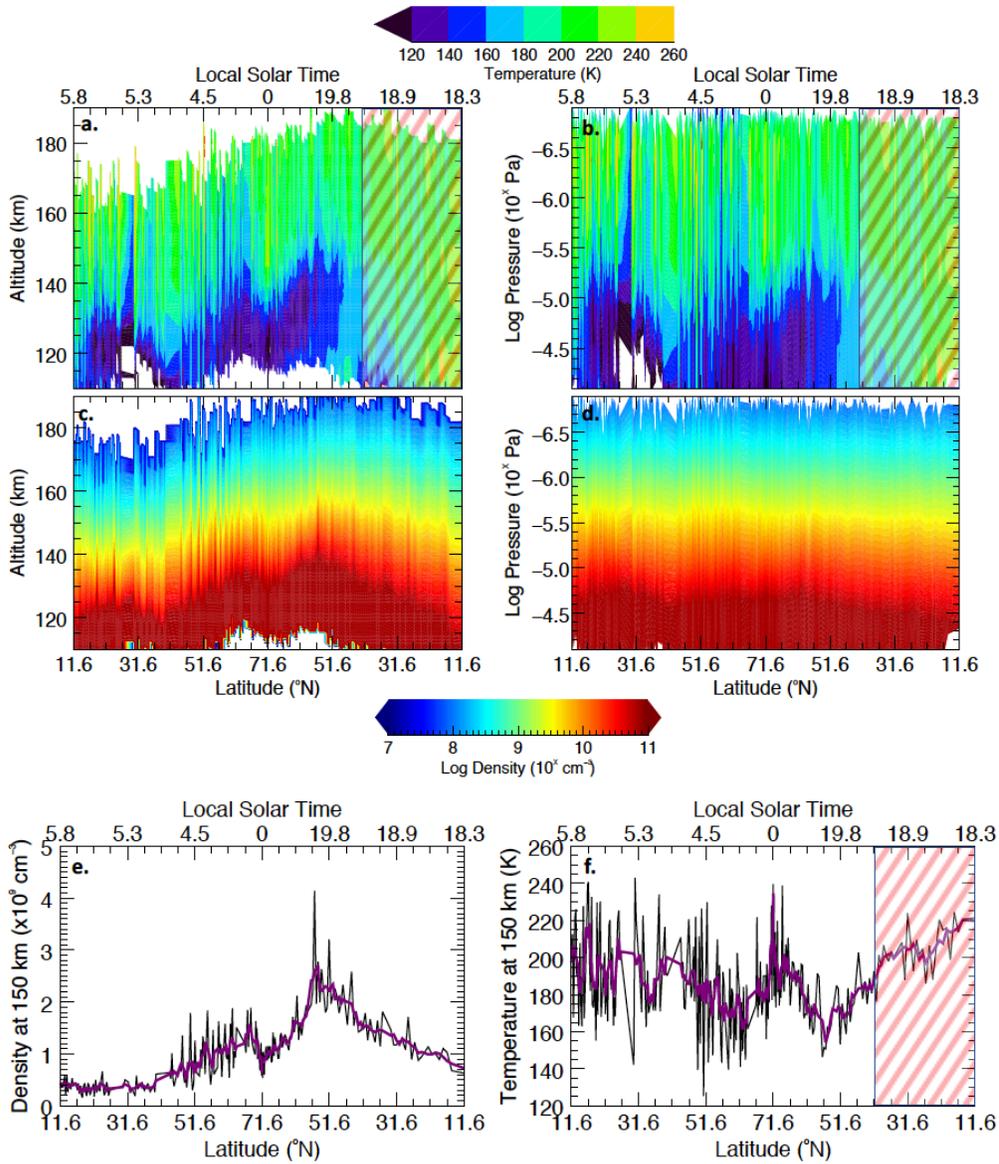

**Figure 6.** Thermospheric temperature and density near aphelion measured by EUVM-SOs. Panels a (c) and b (d) show temperature (density) as a function of latitude and altitude or pressure, respectively. The bottom horizontal axis shows latitude. Ambiguity in latitude between the dawn and dusk terminators can be resolved with the top horizontal axis, which gives the local solar time. Panels e and f show density and temperature at 150 km, with black curves corresponding with individual orbits and purple with the 5-orbit running average.

At a fixed altitude, density increases due to either bulk atmospheric expansion or local heating. At constant pressure, the ideal gas law requires decreasing density to correspond with increasing temperature and presumably increasing heating. As such, plotting density versus pressure controls for changes due to bulk atmospheric expansion (or contraction), isolating local heating (or cooling). Figure 6d shows a decrease in density at $10^{-5}$ Pa moving equatorward on the Dusk Terminator, indicating local heating is increasing away from the North Pole. A temperature maximum is expected due to solar heating at sub-solar latitudes, both directly by increasing EUV heating of the thermosphere, and indirectly by decreasing molecular conduction as result of increased IR heating at lower altitudes [Bougher *et al.*, 1999; 2000]. Figure 6d also shows a decrease in density near 40°N on the dawn-side and Figures 6a and 6b show a corresponding temperature increase, indicating that local heating is occurring here. This feature may be evidence of adiabatic heating from solar driven zonal circulation that is expected to drive upwelling (and adiabatic cooling) in the afternoon sector resulting in subsidence (and adiabatic heating) in the post-midnight sector near sub-solar latitudes, resulting in a "tongue" of increased temperature extending towards dawn [Bougher *et al.*, 1999].

5.2 Mars Dusk Thermospheric Structure near Perihelion

MAVEN was at Mars during the perihelions of Mars Years (MY) 32 and 33, occurring near the end of 2014 and 2016, respectively. EUVM scanned the dusk terminator from 67°S to 30°N between early December 2014 and mid-February 2015, during which the Mars $L_s$ precessed from 244° to 291°, encompassing both perihelion (at $L_s$=251°) and the Northern Winter Solstice (at $L_s$=270°). During MY 33, EUVM scanned from 70°N to 30°N between mid-September 2016 and mid-January 2017, during which the Mars $L_s$ precessed from 223° to 299°. Data from both of these periods are used to construct the maps of temperature and density at the dusk terminator shown in Figure 7, where the panel arrangement is the same as Figure 6, but the LST axis is omitted because all data are at dusk. Data southward of 30°N were measured during MY 32 and the remaining data were measured during MY 33. The two data gaps between 5°N and 30°N coincide with when EUVM was not sun pointed.

Figure 7c shows a large-scale trend of decreasing density at high altitudes from the southern-most latitudes to the northern-most latitudes. This trend disappears in Figure 7d, indicating the southern, summer hemisphere is inflated relative to the northern, winter hemisphere. The high-altitude (exospheric) temperatures are near 250°K from the southern-most latitudes to the low-latitude northern hemisphere. Above 30°N, exospheric temperatures are near 200°K, except for between 52°N and 62°N where the exospheric temperature increases to 260°K. This temperature enhancement is most pronounced above 170 km but extends to the lower thermosphere. This feature is consistent with "polar warming" features predicted to occur at intermediate latitudes during northern winter that result from enhanced inter-hemispheric circulation during perihelion [e.g Bell *et al.*, 2007; Gonzalez-Galindo *et al.*, 2015]. This circulation is enhanced during the dust-season and results in localized adiabatic heating at the North Pole and near 50°N, where the latter is concentrated at higher altitudes. For examples of simulations of this feature and underlying heating rates, see Figures 6 and 7 in Bell *et al.*, 2007 and Figure 11 in Gonzales-Galindo *et al.*, 2009b. Previously, limited "polar warming" features (ΔT ~ 40-60 K) were observed during Odyssey aerobraking (Ls ~ 260-310) only at 120 km at high Northern latitudes (~60-80N) (Bougher et al., 2006; 2017a).

Although not obvious at the resolution shown in Figure 7, a close inspection of the data (not shown here) reveals periodic temperature and density fluctuations with periods of approximately 2 days or, equivalently, 2 degrees in latitude. These may be due to non-migrating tides resulting in enhancements at fixed geographic latitudes that rotate past the terminator while EUVM is making observations. Further analysis is needed to understand the nature of these oscillations and correct for any aliasing caused by the observing cadence.

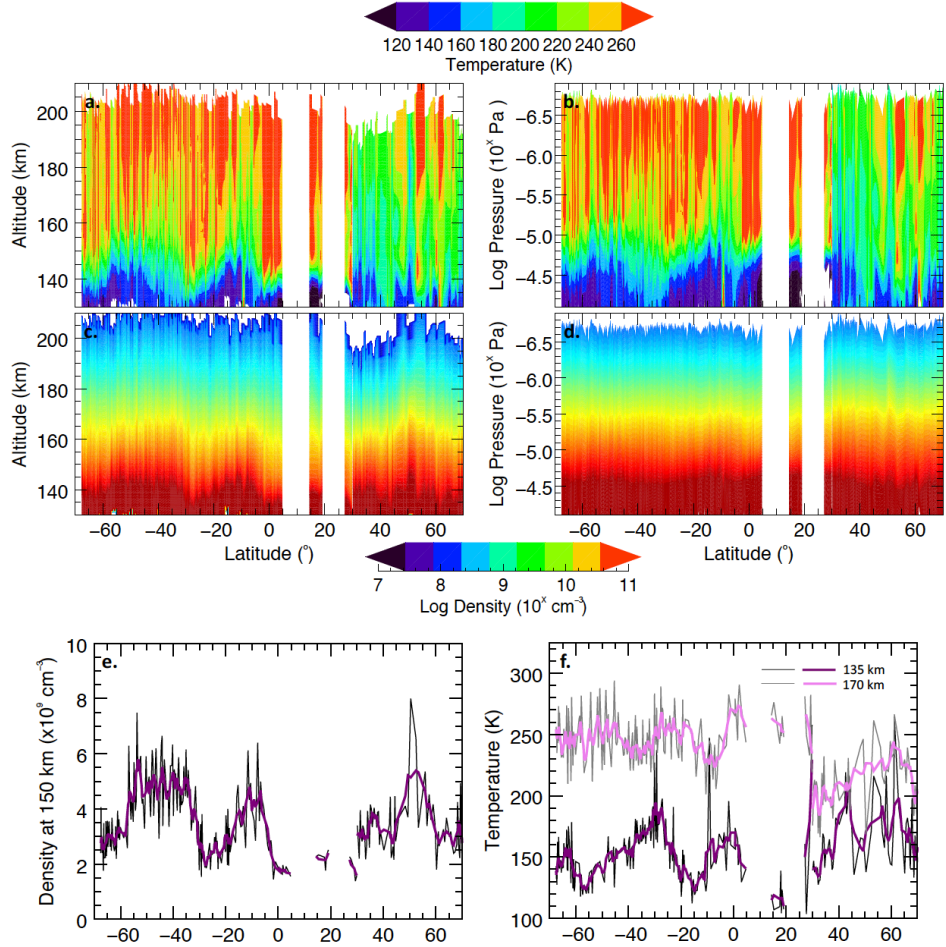

**Figure 7.** Thermospheric temperature and density at the dusk terminator near perihelion measured by EUVM-SOs. Same panel arrangement as Figure 6.

Figure 7c shows two density maxima between 0 and 60°S at 140 km that correspond with pronounced temperature decreases at the same altitude seen in Figure 7a. Although these structures may be latitudinal in nature, there is no obvious correspondence with features arising in published GCM results of the Mars thermosphere near perihelion [Bougher *et al.*, 2000; Gonzales-Galindo *et al.*, 2009b; Bougher *et al.*, 2015a]. Rather, they are believed to be temporal in nature and the result of solar variability because they are co-incident in time with a strong modulation of EUV irradiance due to solar rotation.

Figures 8a and 8c replot the Southern Hemisphere temperature and density data from Figures 7a and 7c, respectively, as functions of time rather than latitude, with the horizontal axis corresponding with days since 19 October 2014. Figure 8b shows the corresponding $E_{EUV}$ from the EUVM Level 3 data product, which shows $E_{EUV}$ varied by 20% with a ~27-day period over this time. Two contours are drawn on Figure 8a to guide the eye to regions of heating and cooling in the thermosphere. Above the black dashed contour, there is a ~27-day heating trend, where temperature increases with increasing EUV irradiance. At lower altitudes, below the dashed white contour, the opposite trend occurs, where temperature decreases with increasing EUV irradiance prior to day ~80. The phase differences can be seen clearly in Figure 7f, which shows temperatures at 135 and 170 km, where the thin curves correspond with individual orbit values and the thicker purple and violet curves correspond with the 5-orbit running average. At 170 km, the running average values vary by ~20 K, which is in excellent agreement with predicted and measured variations in exospheric temperature due to solar rotation variability by Gonzalez-Galindo *et al*., [2015] and Forbes *et al*., [2006], who found values near 25 K and 19.2 K, respectively. The apparent anti-correlation of temperature at 135 km and EUV variability has not been previously predicted nor observed during solar rotations. However, somewhat similar behavior was predicted by Bougher *et al.*, [1999] using MTGCM to simulate the response of the Mars thermosphere to a step-function increase in EUV irradiance, where it was shown that direct heating by EUV irradiance near the altitude of peak EUV heating (~160 km) causes the temperature to increase at high altitudes, causing an expansion of the upper thermosphere. This expansion, in turn, causes the lower thermosphere to cool adiabatically via increasing upward vertical winds, providing a "vertical wind thermostat" that controls the temperature response of the thermosphere due to EUV heating.

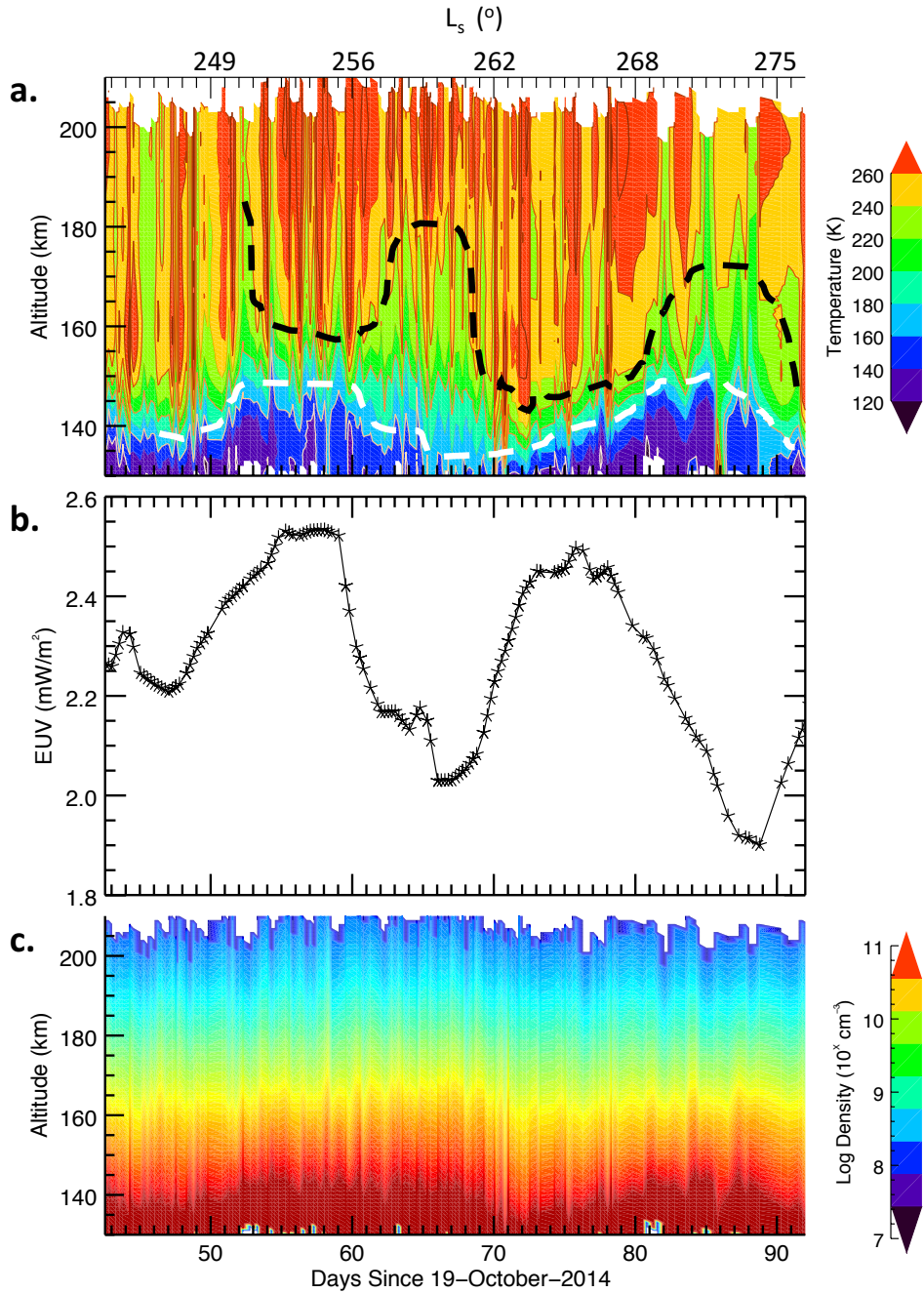

**Figure 8.** Correlation between density and temperature structure and EUV irradiance. a) Thermospheric temperature versus altitude and time. Temperature is correlated with EUV irradiance above the black-dashed contour, but anticorrelated with EUV irradiance below the white-dashed contour. b) Corresponding 0-93 nm integrated EUV irradiance. c) Corresponding thermospheric density.

Features like those appearing below the white-dashed contour in Figure 8a could, in theory, also be caused by regional dust activity that causes an expansion of the lower atmosphere (as a result of increased heating from an elevated opacity), which in turn can lift the mesopause altitude, resulting in a colder, denser atmosphere at a fixed altitude in the lower thermosphere. Dust opacity data for MY 32 is examined to understand the contemporary dust activity. These data are produced using the method of Montabone *et al.* 2015, which uses measurements made by the Thermal Emission Imaging System (THEMIS) instrument onboard Mars Odyssey and the Mars Climate Sounder (MCS) onboard the Mars Reconnaissance Orbiter (MRO), and are publicly distributed on the web at http://www-mars.lmd.jussieu.fr/mars/dust_climatology/. These data are shown in Figure 9a, which shows the zonal mean 15 μm opacity at 610 Pa versus latitude and $L_s$. There is a local maximum opacity of 0.5 near 50$^o$S, which is coincident in latitude with density enhancements and temperature depletions shown in Figure 8. However, this occurred 20 days prior to when EUVM-SOs began for the period shown. By the time EUVM-SO measurements of the Southern Hemisphere began, the opacity at 50$^o$S decayed to 0.25 and gradually increases to a value of 0.35 at the Equator. To evaluate whether this dust activity is expected to result in latent heating capable of causing the structures in lower thermosphere shown in Figure 8, GCM simulations from the Mars Climate Database (MCD) [Forget *et al.*, 1999; Millour *et al.*, 2015; Madeleine *et al.*, 2011] are evaluated for the corresponding times and latitudes using contemporary (MY 32) dust and solar EUV conditions. Figure 9b shows the MCD predictions of temperature between 60 and 200 km, corresponding with the data shown in Figures 7 and 8. The simulations do not predict structures consistent with those observed in Figure 8a, suggesting that they are not a result of latent effects from earlier dust activity at these altitudes, and supporting the conclusion that the observed lower thermospheric cooling is a result of EUV solar rotation variability.

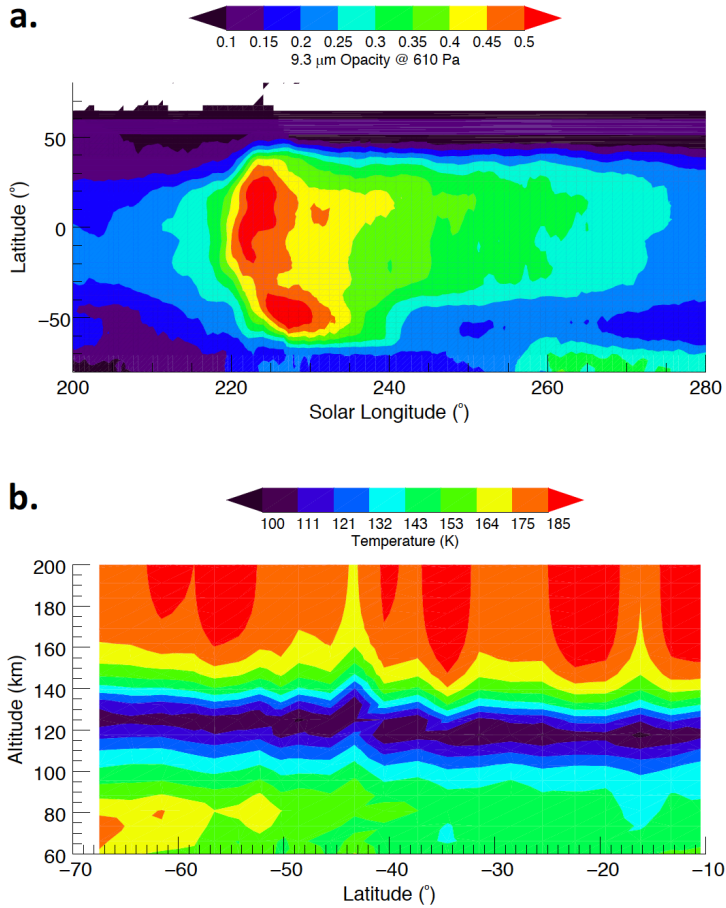

**Figure 9.** MY 32 dust activity and predicted atmospheric response for the time and latitudes of EUVM-SO measurements. a) Distribution of dust versus latitude and solar longitude as measured by 9.3 m vertical column opacity normalized to 610 Pa. b) MCD simulations of temperature near the mesopause corresponding with the data shown in Figure 8.

### 5.3 The Correlation Between Thermospheric Temperature and Solar Variability

Given that the EUVM-SO measurements span two Mars years, during which solar EUV irradiance decreases from solar moderate to solar minimum levels, they provide an opportunity to characterize the dependence of thermospheric temperature on both solar insolation and EUV forcing, where the former is indicative of seasonal forcing on the thermosphere from below and the latter is indicative of direct heating of the thermosphere. Figure 5a shows the thermospheric temperature at 595 nPa, $T_{595}$, along with $E_{EUV}$ and $1/R^2$. The 595 nPa pressure level coincides with approximately 150 km at periapsis and is selected because it sits in between the altitude of expected peak EUV heating near 170 km and peak EUV energy deposition near 130 km [Bougher *et al.*, 1999, Gonzalez-Galindo *et al.*, 2005]. Although $E_{EUV}$ variability reflects the $1/R^2$ variability, the long-term decreasing solar variability dominates, resulting in a 50% reduction of $E_{EUV}$ for the second perihelion as compared to the first perihelion shown. As such, these data should contain adequate variability to decouple the dependence of $T_{595}$ on $E_{EUV}$ and $1/R^2$.

For each terminator, temperatures in four narrow latitude bands are compared with $E_{EUV}$ and $1/R^2$, where the latitude bands are chosen to maximize repeat measurements across the available time range. Figure 10 shows $T_{595}$ and solar variability comparisons for the Dusk Terminator. The left column shows scatterplots of $T_{595}$ versus $E_{EUV}$, where the least-squares best-fit is shown in black, and the right column shows scatterplots of $T_{595}$ versus $1/R^2$. The color-code indicates the time when the measurements were made. Figure 11 is similar to Figure 10 but for the Dawn Terminator. For each panel in Figures 10 and 11, the Pearson correlation coefficient, $r$, is found and reported in Table 1 along with the thermospheric temperature sensitivity to EUV forcing, $dT/dE_{EUV}$. For each geometry, the larger $r$ value is underlined in Table 1 if the difference is statistically significant.

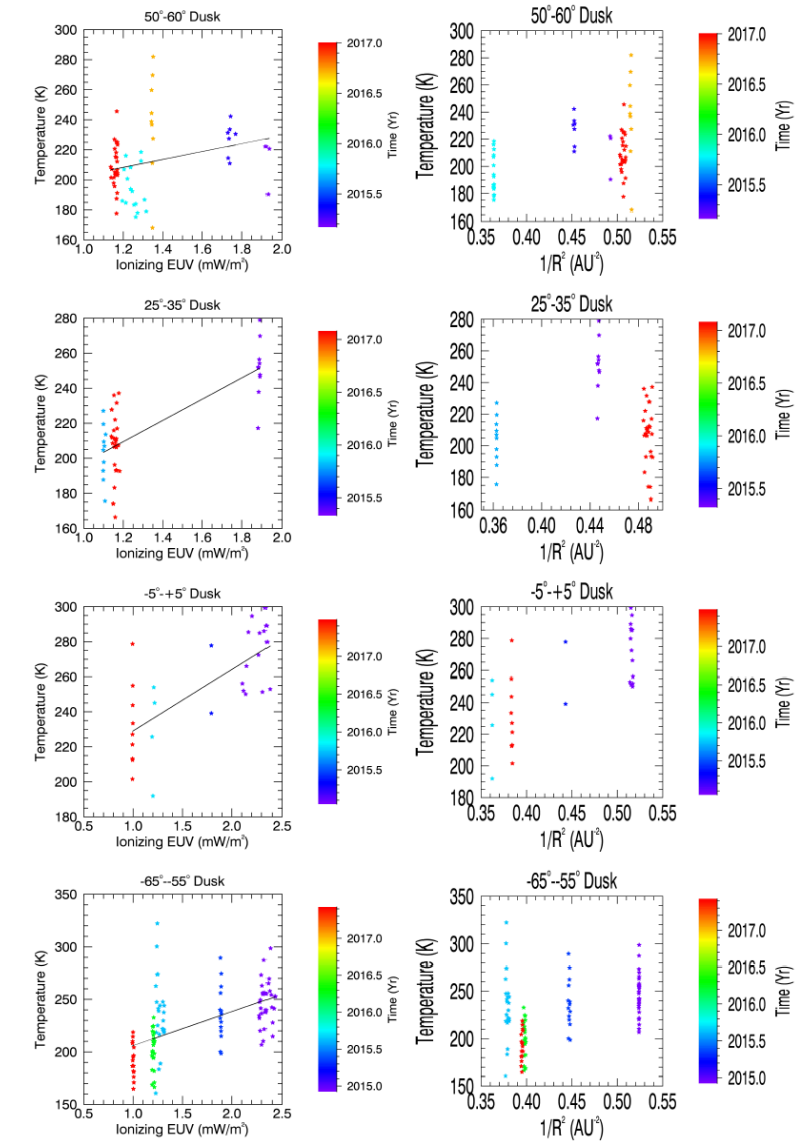

**Figure 10**. Comparisons of thermospheric and solar variability at the Dusk Terminator. The left column shows scatter plots of $T_{595}$ versus $E_{EUV}$ and the right column shows scatterplots of $T_{595}$ versus $1/R^2$. Each row corresponds with a different latitude range.

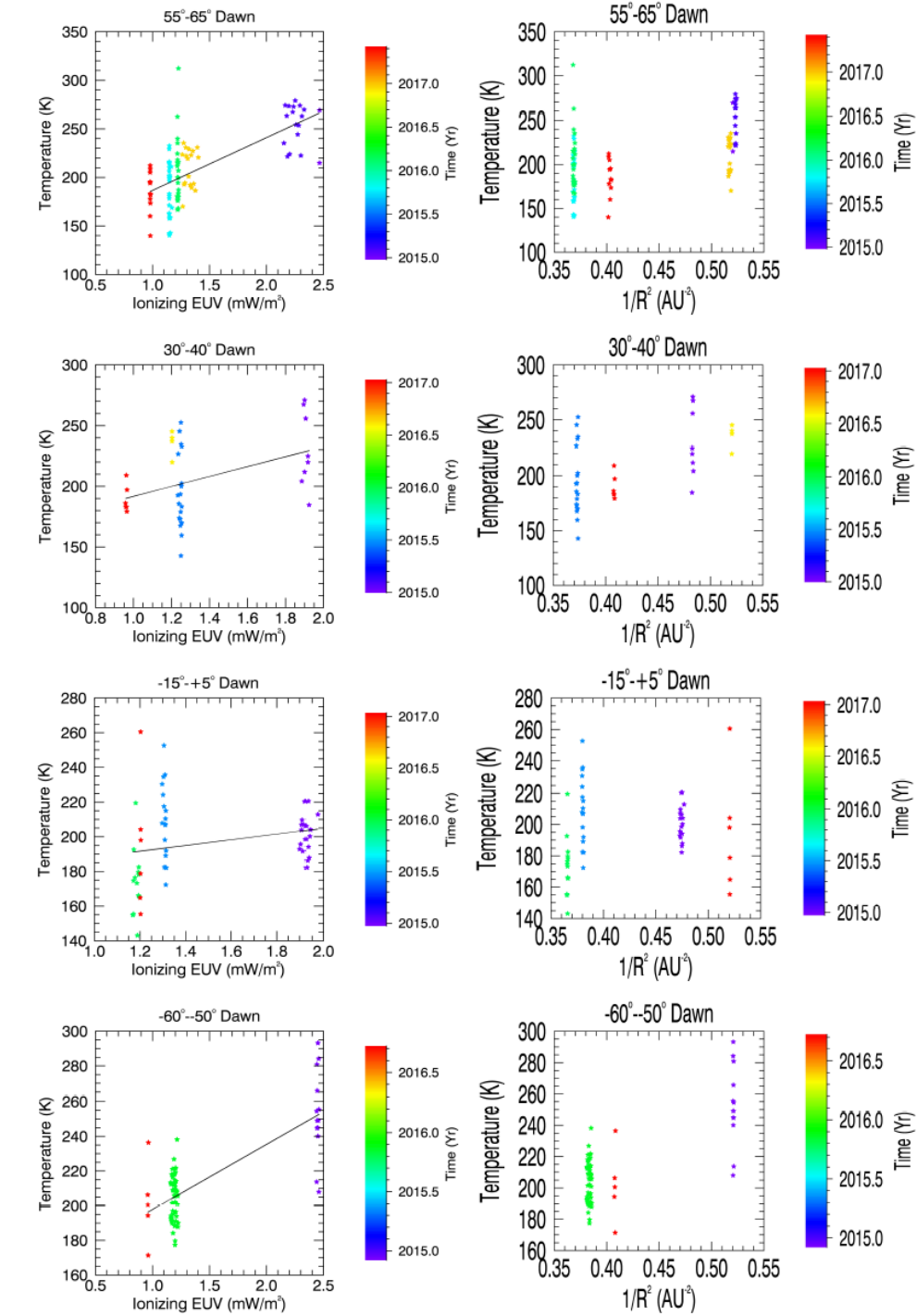

**Figure 11**. Same as Figure 10 but for the dawn terminator.

**Table 1.** Correlation coefficients and temperature sensitivities from the solar and thermospheric variability

comparison shown in Figures 10 and 11.

| LST | Latitude (°) | $r_{EUV}$ | $r_{1/R2}$ | $dT/dE_{EUV}$ (K m²/mW) |
|------|------|------|------|------|
| Dusk | 50- 60 | 0.26 | 0.41 | 26 |
| Dusk | 25- 35 | 0.74 | 0.01 | 61 |
| Dusk | -5- 5 | 0.71 | 0.7 | 35 |
| Dusk | -65- -55 | 0.53 | 0.42 | 32 |
| Dawn | 55- 65 | 0.66 | 0.52 | 54 |
| Dawn | 30- 40 | 0.4 | 0.54 | 41 |
| Dawn | -15- 5 | 0.22 | 0.14 | 16 |
| Dawn | -60- -50 | 0.76 | 0.75 | 38 |

Of the eight geometries considered, $r_{EUV}$ is larger (smaller) than $r_{1/R2}$ for 4 (2) cases; confirming, what is also evident by inspection of Figures 10 and 11: Thermospheric temperature is better correlated with EUV irradiance than solar insolation. This expands the conclusion of Bougher *et al.*, [2017b] who used the MAVEN NGIMS and IUVS instruments to find no significant difference between correlations of thermospheric temperature with $1/R^2$ or $E_{EUV}$. The difference in conclusions is likely due to Bougher *et al.*, [2017b] having used only data between October 2014 and June 2016, a period when the differences between $E_{EUV}$ and $1/R^2$ are relatively small; and the conclusion found here is more complete given the larger extent of data analyzed.

There are some exceptions to better correlation between $T_{595}$ and $E_{EUV}$ found here, most notably, the 50-60°N Dusk case shown in Figure 10. Here, the orange-yellow subset of temperatures are warmer than the trend-line would suggest. These data correspond with when Mars was at perihelion, which is when polar warming was observed at these same latitudes as discussed in Section 5.2. It follows that, although EUV irradiance is the primary energy input to the thermosphere, global circulation plays an important role in determining thermospheric temperature. It is also notable that, while one might expect a higher correlation between $E_{EUV}$ and $T_{595}$ at dusk because the atmosphere there has been heated by the Sun the preceding half-day, the correlation between $E_{EUV}$ and $T_{595}$ is about the same at either terminator, with the average dusk value of $r_{EUV}$ being equal to 0.56 and that at dawn being equal to 0.51.

The average value of $dT/dE_{EUV}$ is found to be 45 ± 12 K m² mW⁻¹ if the two cases where $r_{EUV} < 0.3$ are omitted. $E_{EUV}$ varies from approximately 3.5 mW/m² to 0.75 mW/m² from perihelion at solar maximum to aphelion at solar minimum for a typical solar cycle. A $dT/dE_{EUV}$ value of 45 K m² mW⁻¹ would then correspond with a thermospheric temperature change of 123° K. This value is consistent with the temperature variations predicted by M-GITM at the terminator [Bougher *et al.*, 2015a], but it should be noted that the variability near the sub-solar point is predicted to be about 75° K larger over the solar cycle, reflecting greater solar EUV control of thermospheric temperature here. Gonzalez-Galindo et al., [2015] show a more moderate net variation of ~150°K at the sub-solar point over similar conditions, but do not report temperatures near the terminator. Recently, Bougher *et al.* [2017b] used the MAVEN NGIMS and IUVS instruments to find $dT/dE_{EUV}$ using data from the first Mars Year that MAVEN was at Mars and it found it to be 2.25 km m² mW⁻¹, where the units are in scale-height per solar Lyman-alpha irradiance. Converting this value to the units used here yields 66.7 K m² mW⁻¹. This value is somewhat larger than the value found here, and the differences may be a result of Bougher *et al.*, [2017b] using data with Solar Zenith Angles less than 75° and, therefore, better reflect $dT/dE_{EUV}$ of the day-side, where solar EUV control of thermospheric temperature is greater than it is at the terminator.

## 6 Summary and Conclusions

MAVEN EUVM solar occultations provide a powerful new capability for characterizing the Mars thermosphere and its variability. The channel selected for density retrievals in this study measures wavelengths that are absorbed in the 100-200 km range of the thermosphere. Density retrievals using the other two EUVM channels, which measure irradiance at 0.1-7 nm and 121.6 nm, can potentially extend this range to lower altitudes because photons in these bands penetrate to lower altitudes due to smaller $CO_2$ cross-sections at these wavelengths. As shown in Figure 4, the retrieval uncertainty becomes large at lower altitudes, primarily due to single scale height reference atmospheres being used in the retrieval. As such, multiple scale-height reference atmospheres need to be incorporated into the retrieval method for accurate density retrievals at lower altitudes.

Altitude-latitude maps of thermospheric temperature and density at perihelion and aphelion have been presented, revealing a highly structured and variable thermosphere. Large scale features are generally consistent with the structure revealed by GCMs but a more quantitative and direct comparison is needed in order to both gauge and advance the current understanding of processes that determine the structure of the Mars thermosphere. For example, are the high latitude temperature decreases and density enhancements near aphelion captured in current GCMs? And if not, why? Since gravity wave thermal cooling has been predicted to be substantial at both Earth and Mars [Yiğit, & Medvedev, 2009; Medvedev & Yiğit, 2012], yet never directly observed, understanding whether these high latitude cold features can they be attributed to gravity wave cooling is of significant scientific importance for understanding the thermal balance of both Earth's and Mars's thermospheres. Additionally, a close inspection of the perihelion map reveals periodic temperature and density fluctuations. These may be due to non-migrating tides, but detailed study is needed to understand their origin.

Temperature enhancements from polar warming provide tracers for meridional circulation. Prior limited polar warming observations from accelerometers [Bougher *et al.*, 2006; 2017a] and subsequent GCM studies [Bell *et al.*, 2007] have led to a better understanding of the dynamics of the Mars thermosphere. The EUVM-SO observations provide first-ever vertical profiling of polar warming features at intermediate altitudes, which are in remarkable agreement with GCM predictions. Detailed comparisons of this feature with GCM predictions that account for contemporary dust conditions should lead to improved understanding of meridional circulation, and its variability, at Mars.

Observations of the Mars thermosphere response to solar variability at short timescales, and from solar rotations in particular, provides insight into how thermal balance is maintained in the thermosphere. EUVM-SOs observed the thermal response of both the lower and upper thermosphere to two consecutive solar rotations, providing the first observations of the Mars thermosphere's vertical-wind thermostat, which has been predicted to suppress more extreme fluctuations in dayside thermospheric temperatures due to solar EUV heating [Bougher *et al.*, 1999; 2009]. These observations, coupled with GCM simulations, should lead to a better understanding of the role of adiabatic and 15 μm cooling in maintaining thermospheric temperatures. For example, can the degree of cooling in the lower thermosphere associated with solar rotations be entirely explained by adiabatically induced vertical winds, or does the resulting increased ionization enhance the relative abundance of atomic O in the lower thermosphere to the extent that 15 μm cooling is enhanced?

A broad range of space weather activity has occurred while MAVEN has been at Mars [Lee *et al.*, 2017]. Therefore, the EUVM-SO dataset can be used to study the Mars thermosphere response to transient space weather events including solar flares, coronal mass ejections (CMEs) or solar energetic particles (SEPs). To date, only solar flares have been shown to cause temperature enhancements in the thermosphere [Thiemann *et al*. 2015, Jain *et al* 2018], which have been shown to occur primarily at higher altitudes, be fleeting in nature and increase the exosphere temperature by up to 50-100 $^\circ$K. The enhancement of EUV irradiance during large flares is comparable to the enhancement occurring over a solar cycle [Woods *et al.*, 2006]. It follows, based on the $dT/dE_{EUV}$ estimates in Section 5.3, that enhancements approaching 123$^\circ$K could occur near the terminator if we assume these equilibrium estimates apply over the short time-scales of flares. The extended altitude range afforded by the EUVM-SO could lead to a better understanding as to why flare-induced temperature enhancements occur primarily at high altitudes. Indeed, the EUVM-SO observations of the Mars thermosphere response to a solar rotation may provide an important clue into how thermostatic processes dampen the response of the thermosphere to rapid changes in EUV forcing. Further, previous observations of thermospheric flare response have been limited to when MAVEN periapsis has been on the day-side, which excludes a large portion of the MAVEN mission [Thiemann *et al*. 2015]. As such, the addition of the new EUVM-SO measurements increase the statistical likelihood of observing any neutral response to solar flares as well as CMEs and SEPs.

The EUVM-SO results along with recent results using similar instrumentation and methods at Earth by Thiemann *et al*., [2017a] demonstrate that solar EUV foil filter photometers can effectively characterize a region of planetary thermospheres that has been historically difficult to measure. The simplicity of the instrumentation should be emphasized: The EUVM channel used here consists of a foil filter, limiting aperture, Si diode and low current electrometer. These technologies have substantially matured over the past two decades, allowing for foil filter photometers to be built at relatively low cost and small size, with low power and data requirements. As such, they should be considered by mission planners for future planetary mission flights-of-opportunity as either small sun-pointed sensors or separate 2U sun-pointed CubeSats.

**Acknowledgments and Data**

http://phidrates.space.swri.edu.

The dust opacity maps were downloaded from at http://www-mars.lmd.jussieu.fr/mars/dust_climatology/ and the Mars Climate Database outputs were taken from v5.3 of the web interface available at http://www-mars.lmd.jussieu.fr/mcd_python/; both websites are hosted by the Laboratoire de Meteorlogie Dynamique,

E. T., F. G., M. P., L. A., B. T., M. D., S. B., and B. J. were supported by the NASA MAVEN Program. M. D. was supported by the Belgian Federal Science Policy Office through the ESA-PRODEX programme. Z. G. was supported by an appointment to the NASA Postdoctoral Program at NASA Goddard Space Flight Center, administered by Universities Space Research Association under contract with NASA.

E.T. would like to thank Dr. Sonal Jain of the University of Colorado, Boulder; Dr. R. Lillis of the University of California, Berkeley; and Dr. H. Groller of the University of Arizona, Tucson for their helpful conversations on the topic of this study. E.T. also thanks Dr. M. West of the Royal Observatory of Belgium for assistance with the PROBA2 SWAP images.